\documentclass[journal,10pt]{IEEEtran}

\usepackage{amsmath,amssymb,amsfonts}
\usepackage{tabularx}
\usepackage{color}
\usepackage{cite}
\usepackage{algorithm,algorithmic}
\usepackage{pstool}
\usepackage{psfrag}
\usepackage{lettrine}
\usepackage{float}
\usepackage[export]{adjustbox}
\usepackage[caption = false]{subfig}
\usepackage{graphicx}
\usepackage{booktabs}
\usepackage[dvips]{epsfig}

\DeclareMathOperator*{\argmax}{argmax}
\begin{document}
%
\title{Adaptive Beamwidth Control for \\ mmWave Beam Tracking
\thanks{Hyeonjin Chung and Sunwoo Kim are with the Department of Electronics and Computer Engineering, Hanyang University, Seoul, 04763, South Korea (email: hyeonjingo@hanyang.ac.kr; remero@hanyang.ac.kr).}}

\author{
\IEEEauthorblockN{Hyeonjin Chung,~\IEEEmembership{Student Member,~IEEE,} and Sunwoo Kim,~\IEEEmembership{Senior Member,~IEEE}\\}}
\maketitle

\begin{abstract}
Traditional beam tracking methods have severe performance loss under the high mobility and narrow beam scenario. To alleviate the tracking performance degradation, we propose an adaptive beamwidth control for millimeter wave (mmWave) beam tracking. The particle filter is applied to the beam tracking, and the AoA estimation error is approximated with a posterior density function constructed by the particles. The error approximation leads to the adaptive beamwidth control which is implemented by the partial activation of the antenna array. 
Simulation results show that the proposed algorithm aids the beam tracking to yield a smaller AoA estimation error under the high mobility environments.
\end{abstract}


\IEEEpeerreviewmaketitle

\section{Introduction}
\lettrine[
  loversize=0.2,
  lines=2,
  findent=0.2em,
  nindent=0.03em,]{\textbf{O}}{NE} of the key enablers of millimeter wave (mmWave) communication is beamforming \cite{6736750}. Due to the high directivity, the beam between a base station (BS) and a user equipment (UE) must be aligned. However, the misalignment of the beam may occur when the UE is mobile. The beam tracking methods were proposed to solve the misalignment by updating the beam steering direction toward the signal~\cite{7905941,8025577}. They are expected to suffer from the high mobility of the UE due to the narrow beam. Therefore, the method to alleviate the performance degradation induced by the high mobility is crucial for a seamless mmWave beam tracking.    

The beamwidth control schemes which can alleviate the influence of the high mobility were proposed in \cite{BWA_60ghz,BWA_Design, BWA_Beamsweeping}. For the higher mobility, the transmitter radiates a wider beam which is pre-set in the codebook~\cite{BWA_60ghz}. 
In~\cite{BWA_60ghz}, it assumes that the mobility of the UE is known, which is usually unavailable in practice.
Also, the beamwidth control under a vehicular environment was proposed in \cite{BWA_Design, BWA_Beamsweeping} which allocate the optimal beamwidth for each section on the road based on given velocity of the UE. However, these schemes are not suitable for the Bayesian tracking framework. 
    
In this paper, we propose an adaptive beamwidth control for mmWave beam tracking, where the mobility of the UE is unknown. The particle filter-based beam tracking updates the beam steering angle, and the AoA estimation error is approximated by a posterior distribution generated by the particles. Based on the error approximation, the beamwidth is adaptively adjusted by the partial activation of the antenna array.
The simulation results compare the proposed algorithm with the existing particle filter-based tracking~\cite{7511857} to confirm the performance improvement under the high mobility of the UE.  

\section{Signal and Channel Model}\label{system model}
We assume that the uniform linear array (ULA) is used on the BS while one omni-directional antenna is used on the UE. For an uplink scenario, the BS and the UE serve as a receiver and a transmitter. It is known as a single-input-multiple-output (SIMO) system. The steering vector of the ULA, $\mathbf{a}(\phi,M)$, is
\begin{equation}
\begin{split}
\label{steer_vec}
    \mathbf{a}(\phi,M) = [1,e^{j\pi\cos\phi},\ldots,e^{j\pi(M-1)\cos\phi}]^T \in \mathbb{C}^{M\times 1},
\end{split}
\end{equation}
where $\phi$ denotes the AoA from UE, and $M$ is the number of active antennas, which is an argument controlling the beamwidth in the BS. The spacing between adjacent antennas is set to half wavelength. 

The SIMO channel at the $k$-th time slot, $\mathbf{h}_{k}$, can be given as
\begin{equation}\label{channel_model}
    \mathbf{h}_k=\sum_{l=1}^{L}\alpha_{k}^{l}\mathbf{a}(\phi_{k}^{l},{M}_{k-1}) \in \mathbb{C}^{{M}_{k-1} \times 1},
\end{equation}
where $L$ is the number of paths, $\alpha_{k}^{l}$ and $\phi_{k}^{l}$ respectively are the channel gain and the AoA of the $l$-th path, and ${M}_{k-1}$ is the number of active antennas determined at the $(k-1)$-th time slot by the adaptive beamwidth control.

The received signal is discretely sampled to yield the $d$-th sample at the $k$-th time slot, ${z}_k[d]$, where
\begin{equation}\label{rx_signal_one_sample}
\begin{split}
    {z}_k[d] =\mathbf{w}(\hat{\phi}_{k-1},{M}_{k-1})^{H} \mathbf{h}_k {q}[d]+\mathbf{w}(\hat{\phi}_{k-1},{M}_{k-1})^{H}\mathbf{n}.
\end{split}
\end{equation}
$\mathbf{w}(\hat{\phi}_{k-1},{M}_{k-1})$ is the receive beamforming vector with $M_{k-1}$ active antennas, and its direction is set to $\hat{\phi}_{k-1}$ which is the steering angle of the beam determined by the beam tracking procedure at the $(k-1)$-th time slot.
$q[d]$ is the $d$-th sample of the unit-energy pilot signal and $\mathbf{n}\sim \mathcal{CN}(\mathbf{0},N_{0}\mathbf{I})$.  
Letting the $D$ measurement samples are used for tracking, 
the received signal at the $k$-th time slot, $\mathbf{z}_{k}$, is
\begin{equation}\label{rx_signal}
\begin{split}
    \mathbf{z}_k &=[{z}_{k}[1],{z}_{k}[2],\ldots,{z}_{k}[D]]\\&=\mathbf{w}(\hat{\phi}_{k-1},{M}_{k-1})^{H} \mathbf{h}_k \mathbf{q} + \mathbf{w}(\hat{\phi}_{k-1},{M}_{k-1})^{H}\mathbf{N} \in \mathbb{C}^{1 \times D},
\end{split}
\end{equation}
where $\mathbf{q} = [q[1], q[2], \ldots, q[D]] \in \mathbb{C}^{1 \times D}$, and satisfies such that $\mathbb{E}[\mathbf{q}\mathbf{q}^{H}]=1$. $\mathbf{N} \in \mathbb{C}^{{M}_{k-1} \times D}$ is a noise matrix which is a set of independent noise vectors.
Assuming the use of a massive array, an analog beamformer consisting solely of a phase shifter is used when the practical issues are considered \cite{5447703}. Under this assumption, $\mathbf{w}(\hat{\phi}_{k-1},{M}_{k-1})=\mathbf{a}(\hat{\phi}_{k-1},{M}_{k-1})$.

\begin{figure}[!t]
    \psfrag{a}[Bc][bc][0.7][0]{Analog to digital converter}
	\psfrag{b}[Bc][bc][0.7][0]{Antenna}
	\psfrag{c}[Bc][bc][0.7][0]{Activated}
	\psfrag{o}[Bc][bc][0.7][0]{Deactivated}
	\psfrag{d}[Bc][bc][0.7][0]{Switch}
	\psfrag{e}[Bc][bc][0.7][0]{Phase shifter}
	\psfrag{s}[Br][bc][0.7][0]{Combiner}
	\psfrag{g}[Bc][bc][0.6][0]{$1$}
	\psfrag{j}[Bc][bc][0.6][0]{$2$}
	\psfrag{n}[Bc][bc][0.6][0]{$3$}
	\psfrag{y}[Bc][bc][0.6][0]{$M_k$}
	\psfrag{i}[Bc][bc][0.6][0]{$M_k+1$}
	\psfrag{i5}[Bc][bc][0.6][0]{$M_k+2$}
	\psfrag{i6}[Bc][bc][0.6][0]{$M_0$}
    \includegraphics[width=0.93\columnwidth,right]{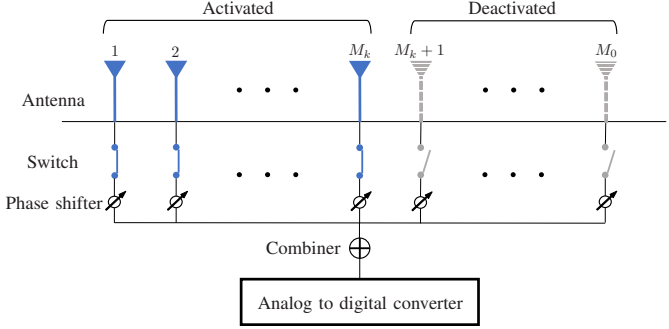}
    \caption{The control of the antenna array for the beamwidth adjustment.}
    \label{array}
\end{figure}

Since the beam is steered to the direction of the line-of-sight (LOS) signal from the UE, and the non-line-of-sight (NLOS) signal tends to be 10-20 dB weaker than the LOS signal~\cite{6387266}, we assume that the signals from the NLOS paths can be neglected. Then, $\mathbf{z}_{k}$ can be approximated as
\begin{align}\label{rx_signal2}
\begin{split}
    \mathbf{z}_{k} \approx
    &\underbrace{\alpha_{k} \mathbf{w}(\hat{\phi}_{k-1},{M}_{k-1})^{H} \mathbf{a}(\phi_{k-1},{M}_{k-1}) \mathbf{q}}_{\textrm{Signal}} +\\ &\underbrace{\mathbf{w}(\hat{\phi}_{k-1},{M}_{k-1})^{H}\mathbf{N}}_{\textrm{Noise}},
\end{split}
\end{align}
where $\alpha_k=\alpha_{k}^{1}$ and $\phi_k=\phi_{k}^{1}$, which denote the channel gain and the AoA of the LOS signal. 

At the $k$-th time slot, the beam tracking and the adaptive beamwidth control are executed based on $\mathbf{z}_{k}$ to update $\hat{\phi}_{k}$ and ${M}_{k}$. For the beam tracking, the particle filter is used to obtain $\hat{\phi}_{k}$ which is the estimation of $\phi_{k}$. Then, the beamwidth is adjusted by the partial activation of the array as shown in Fig. \ref{array}. The number of active antennas cannot exceed the total number of antennas $M_{0}$, and only the antennas on the edge of the array can be deactivated to maintain the equal spacing between adjacent antennas. For a large AoA estimation error, the beamwidth is widened to mitigate the drop of the received signal strength induced by the beam misalignment.   
Since the sharpness of the beam is proportional to the number of active antennas, the proposed beamwidth control selects a smaller value for ${M}_{k}$ as the AoA estimation error becomes larger.

By using (\ref{rx_signal2}), the receive SNR obtained after the beam tracking and the adaptive beamwidth control at the $k$-th time slot can be denoted as
\begin{equation}\label{SNR_SIMO}
\begin{split}
    \gamma_{k}&=\frac{\mathbb{E}[( \alpha_{k} \mathbf{w}_{k}^{H} \mathbf{a}_{k} \mathbf{q} )( \alpha_{k} \mathbf{w}_{k}^{H} \mathbf{a}_{k} \mathbf{q})^{H}]}{\mathbb{E}[( \mathbf{w}_{k}^{H} \mathbf{N})  (\mathbf{w}_{k}^{H} \mathbf{N})^{H}]}\\&= \frac{\left\vert \alpha_{k} \right\vert^2} {N_{0}} \frac{ \mathbf{w}_{k}^{H} \mathbf{a}_{k} \mathbf{a}_{k}^{H} \mathbf{w}_{k}}{\mathbf{w}_{k}^{H}\mathbf{w}_{k}}=\frac{\left\vert \alpha_{k} \right\vert^2}{N_{0}} \frac{\left\vert \mathbf{w}_{k}^{H} \mathbf{a}_{k} \right\vert^{2}}{{M}_{k}},
\end{split}
\end{equation}
where $\mathbf{w}(\hat{\phi}_{k},{M}_{k})$ and $\mathbf{a}(\phi_{k},{M}_{k})$ are respectively shortened to $\mathbf{w}_{k}$ and $\mathbf{a}_{k}$. Excluding the constants in (\ref{SNR_SIMO}), $\gamma_{k}$ is dependent to $\hat{\phi}_{k}$ and ${M}_{k}$, which are the outputs of the beam tracking and the adaptive beamwidth control.

\section{The Proposed Adaptive Beamwidth Control for mmWave Beam Tracking}\label{BWA_AIDED}
\subsection{Particle Filter-based Beam Tracking}\label{PFTracking}
The state model for the particle filter-based beam tracking, $\mathbf{x}_{k}=[\alpha^R_{k},\alpha^I_{k},\phi_{k}]^{T}$, 
where $\alpha^R_{k}$ and $\alpha^I_{k}$ respectively are the real and the imaginary part of the channel gain $\alpha_k$. 
Discrete time stochastic evolution model is given by $\mathbf{x}_{k}=\mathbf{F}\mathbf{x}_{k-1}+\mathbf{u}$, where $\mathbf{F} \approx \mathbf{I}$, assuming the period between adjacent time slots is sufficiently small~\cite{7511857}. The process noise, $\mathbf{u}\sim\mathcal{N}(\mathbf{0},\mathbf{\Sigma_{u}})$, where $\mathbf{\Sigma_u}=\textrm{diag}[\sigma_{\alpha^{R}}^{2},\sigma_{\alpha^{I}}^{2},\sigma^2_{\phi}]$. Since $\mathbf{u}$ denotes the variation of the channel from the prior time slot, it grows in proportion to the mobility of the UE.   

The particle filter generates $S$ random particles as
\begin{equation}\label{propagation}
    \mathbf{x}_{k}^{s}\sim \mathcal{N}(\mathbf{F}\mathbf{x}_{k-1}^{s},\mathbf{\Sigma_{u}}), \; \textrm{for} \; s=1,\ldots,S,
\end{equation}
where $\mathbf{x}_{k}^{s}$ denotes the $s$-th particle. The weights of the particles are given by 
\begin{equation}
    w_{k}^{s} = w_{k-1}^{s} \mathcal{L}(\mathbf{z}_{k}|\mathbf{x}_{k}^{s}),\; \textrm{for} \; s=1,\ldots,S,    
\end{equation}
where $\mathcal{L}(\mathbf{z}_{k}|\mathbf{x}_{k}^{s})$ is the likelihood of $\mathbf{z}_{k}$ given $\mathbf{x}_{k}^{s}$. The weights are then normalized to satisfy such that $\sum_{s=1}^{S} w_{k}^{s}=1$.
For ensuring the diversity of the particles, they undergo resampling step~\cite{PF}.  
After the resampling, the estimated state can be yielded as 
\begin{equation}
    \hat{\mathbf{x}}_{k}= \sum_{s=1}^{S}{w_{k}^{s} \mathbf{x}_{k}^{s}},
\end{equation}
where the steering angle of the beam is updated to $\hat{\phi}_{k}$.

\begin{figure}[!t]
    \centering
    \psfrag{a}[Br][bc][0.7][0]{UE}
	\psfrag{e}[Br][bc][0.7][0]{2. AoA error}
	\psfrag{v}[Bc][bc][0.7][0]{$\mathbf{x}_{k-1}$}
	\psfrag{m1}[Bc][bc][0.7][0]{$M_{k-1}$}
	\psfrag{m2}[Bc][bc][0.7][0]{$M_{k}$}
	\psfrag{c}[Bc][bc][0.7][0]{$\mathbf{x}_{k}$}
	\psfrag{o}[Bc][bc][0.7][0]{$\hat{\mathbf{x}}_{k}$}
	\psfrag{b}[Bc][bc][0.7][0]{3. Adaptive beamwidth control}
	\psfrag{f}[Br][bc][0.7][0]{BS}
	\psfrag{g}[Bc][bc][0.7][0]{\shortstack[l]{1. Estimate AoA using particle filter}}
    \includegraphics[width=0.94\columnwidth]{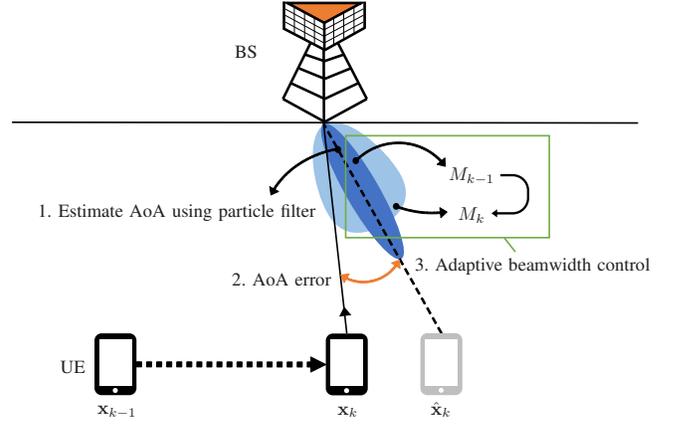}
    \caption{The schema of the beamwidth control-aided tracking.}
    \label{BWA+tracking}
\end{figure}


\subsection{Adaptive Beamwidth Control}\label{ACB}

When the AoA estimate deviates from the true signal direction as much as the signal cannot be captured within the beamwidth,
the received signal strength will significantly drop yielding the communication failure.
The proposed beamwidth control can prevent such failure by adaptively widening or narrowing the beamwidth as shown in Fig. \ref{BWA+tracking}. The beamwidth is determined by the degree of the AoA estimation error which is explained as follows.

The mean squared error of the AoA estimate at the $k$-th time slot given the measurement $\mathbf{z}_{k}$ is represented and approximated as 
\begin{equation}\label{cond_var}
\begin{split}
\mathbb{E}[(\phi_{k}-\hat{\phi}_{k})^{2}|\mathbf{z}_{k}] &= \int{(\phi_{k}-\hat{\phi}_{k})^{2}p(\phi_{k}|\mathbf{z}_{k})}d\phi_{k} \\
       &\approx \sum_{s=1}^{S}{w_{k}^{s}(\phi_{k}^{s}-\hat{\phi}_{k})^{2}}.
\end{split}
\end{equation}
The approximation is possible with the particle samples that are used in the tracking algorithm~\cite{PF}.
We define the root mean square error (RMSE) of the AoA estimate as
\begin{equation}\label{app_part}
   e_{k}=\sqrt{\sum_{s=1}^{S}{w_{k}^{s}(\phi_{k}^{s}-\hat{\phi}_{k})^{2}}},
\end{equation}
as it will be used in the following derivation.



$M_{k}$ which denotes the beamwidth at the $k$-th time slot is obtained such that it maximizes the receive SNR $\gamma_{k}$. Thus,
\begin{equation}\label{Object_Func}
M_k =\argmax_{M}\frac{| \mathbf{w}(\hat{\phi}_{k},M)^{H} \mathbf{a}({\phi}_{k},M) |}{\sqrt{M}}, \; \textrm{s.t.} \; M >0.
\end{equation}
As shown in Appendix \ref{app1}, the maximization can be be reformulated as follows.
\begin{equation}\label{Object_Func2}
\begin{split}
M_{k}=\argmax_{M}\left|\frac{\sin (\pi M (\cos \hat{\phi}_{k}-\cos \phi_{k})/2)}{\sqrt{M}\sin (\pi (\cos \hat{\phi}_{k}-\cos \phi_{k})/2)}\right| , & \\ \textrm{s.t.} \; M > 0.&
\end{split}
\end{equation}
The solution for (\ref{Object_Func2}) is given as
\begin{equation}\label{result}
{M}_{k}=\frac{2.330}{\pi \left| \cos\hat{\phi}_{k}-\cos{\phi_{k}} \right|},
\end{equation}
where its derivation is shown in Appendix \ref{app2}.   

Note that $M_{k}$ cannot be calculated as shown in (\ref{result}) since the true $\phi_{k}$ is unknown in practice. We assume that $\phi_{k}$ is approximated as $\hat{\phi}_{k}+e_{k}$ or $\hat{\phi}_{k}-e_{k}$ with an equal probability.
Then, $M_{k}$ is now
\begin{equation}\label{final}
\begin{split}
    &{M}_{k}=\\&\left\lfloor {\frac{1.165/\pi}{\left\vert \cos(\hat{\phi}_{k})-\cos(\hat{\phi}_{k}+e_{k}) \right\vert}+ \frac{1.165/\pi}{\left\vert \cos(\hat{\phi}_{k})-\cos(\hat{\phi}_{k}-e_{k}) \right\vert}}\right\rceil.
\end{split}
\end{equation}
The rounding is applied in (\ref{final}) to make $M_k$ an integer value, where $\left\lfloor(\cdot)\right\rceil$ denotes the rounding of the argument.
For a small $e_{k}$, ${M}_{k}$ may become larger than $M_{0}$, then the entire array elements will be activated to have the finest beamwidth.

\section{Simulation Results and Discussion}\label{simulation}
\begin{table}[!t]
  \begin{center}
    \caption{Simulation parameters}
    \label{tab:per}
    \begin{tabular}{c|c|c|c|c|c}
    \toprule
     Parameter &$\alpha_0$ & $\phi_0$ & $\sigma_{\alpha}^{R},\sigma_{\alpha}^{I}$ & $\sigma_{\phi}$ & $10 \log\gamma_{0}$  \\ \midrule
     Value&$(1+j)/{\sqrt{2}}$ & $90^{\circ}$ & 0.1 & $1.0^{\circ}$ & 20 dB  \\
     \bottomrule
  \end{tabular}
   \end{center}
\end{table}
We compared the performance of the proposed algorithm with the particle filter-based tracking without the adaptive beamwidth control~\cite{7511857}. SIMO system is considered, and the BS receives the signal with 64 antennas. The number of samples used for the tracking, $D$, and the number of the particles, $S$, are set to 5 and 1000 respectively.
To simulate a high mobility of the UE, the parameters which dictate the variation of the channel, $\sigma_{\alpha}^{R}$, $\sigma_{\alpha}^{I}$, and $\sigma_{\phi}$, are set larger than those in~\cite{7905941,8025577,7511857}. The other parameters are given in Table \ref{tab:per}.
 

\begin{figure}[!t]
    \centering
    \includegraphics[width=\columnwidth]{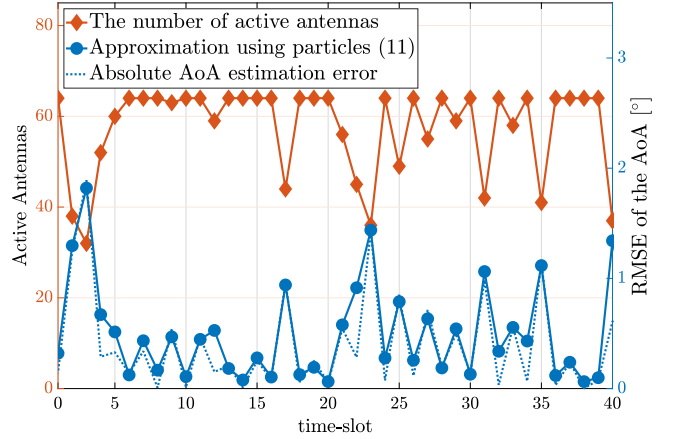}
    \caption{The correlation between the number of active antennas, the approximation of (\ref{app_part}), and the absolute AoA estimation error during the beamwidth control-aided tracking.}
    \label{tracking}
\end{figure}
As shown in Fig. \ref{tracking}, the number of active antennas is inversely proportional to the AoA estimation error. For the beamwidth widening, more antennas are deactivated as the AoA estimation error increases, whereas antennas up to 64 are activated for the small error to reconstruct the fine beam. Also, it is shown that the approximation of (\ref{app_part}), which is drawn from the particles has a strong correlation with the absolute AoA estimation error.

\begin{figure}[!t]
    \centering
    \includegraphics[width=\columnwidth]{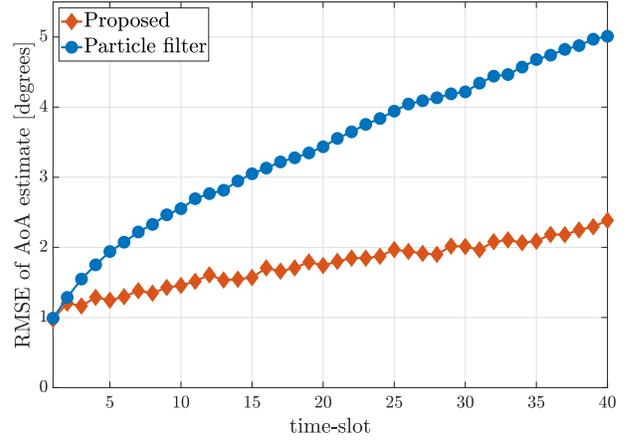}
    \caption{RMSE results of the beamwidth control-aided tracking and the particle filter-based beam tracking. A total of 1000 Monte Carlo runs are used for the RMSE calculation.}
    \label{RMSE}
\end{figure}

Fig. \ref{RMSE} shows the RMSE of the AoA estimate over the time. It is clearly demonstrated that the proposed algorithm outperforms the particle filter-based tracking without the adaptive beamwidth control in terms of RMSE.
The result confirms that the adaptive beamwidth control successfully improves the AoA tracking under the high mobility of the UE. 

However, the chance of the tracking failure still exists due to the excessive AoA variation, in which the signal direction goes beyond the current beamwidth.
Once the beam fails to capture the signal from the UE, the estimation error starts to diverge since there is no received signal to use for the state estimation.
In this case, it is possible that the particle error approximation can be poor, which leads to the overall performance degradation. Still, the proposed algorithm is an effective solution for the beam tracking in the high mobility scenarios.


\section{Conclusion}
In this paper, we have proposed an adaptive beamwidth control which complements the mmWave beam tracking. The particle filter is applied for the beam tracking, and the AoA estimation error is approximated by a posterior distribution derived from the particles. The beamwidth control is implemented by the partial activation of the antenna array, and the optimal number of active antennas is derived by maximizing the receive SNR. 
Simulation results confirm that the proposed algorithm presents a notable tracking improvement compared to the existing particle filter-based beam tracking under the high mobility. 

\begin{appendices}
\section{Derivation of (\ref{Object_Func2})} \label{app1}
The objective function of (\ref{Object_Func}) can be expressed as the discrete time Fourier transform (DTFT).
\begin{equation}
\begin{split}
    \label{DTFT1}
    \frac{\left\vert \mathbf{w}(\hat{\phi}_{k},M)^{H} \mathbf{a}({\phi}_{k},M) \right\vert}{\sqrt{M}}&= \left|\sum_{m=0}^{M-1} \frac{1}{\sqrt{M}} e^{-j m\pi (\cos \hat{\phi}_{k}-\cos \phi_{k})}  \right|\\&=\left|\sum_{m=0}^{M-1} \frac{1}{\sqrt{M}} e^{-j \Omega m}  \right|,
\end{split}
\end{equation}
where $\Omega=\pi(\cos \hat{\phi}_{k}-\cos \phi_{k})$.
The problem corresponds to applying DTFT to a rectangular pulse with $M$ samples, whose amplitude is $1/\sqrt{M}$. Thus, the following equalities can be obtained. 
\begin{equation}
\begin{split}
\left|\sum_{m=0}^{M-1} \frac{1}{\sqrt{M}} e^{-j \Omega m}\right|&=
\left|\frac{\sin (M \Omega/2)}{\sqrt{M}\sin(\Omega/2)}\right| \quad \quad \\&=
\left|\frac{\sin (\pi M (\cos \hat{\phi}_{k}-\cos \phi_{k})/2)}{\sqrt{M}\sin (\pi (\cos \hat{\phi}_{k}-\cos \phi_{k})/2)}\right|.
\end{split}
\end{equation}

\section{Derivation of (\ref{result})} \label{app2}
We define $c(M)$ as the objective function of (\ref{Object_Func2}) as follows. 
\begin{equation} 
  c(M)=\left|\frac{\sin (\pi M (\cos \hat{\phi}_{k}-\cos \phi_{k})/2)}{\sqrt{M}\sin (\pi (\cos \hat{\phi}_{k}-\cos \phi_{k})/2)}\right|.
\end{equation}
Finding the local maximum points of $c(M)$ can be represented as   
\begin{equation}
\frac{d c(M)}{dM}=\frac{1}{\sin (\Omega/2)} \left[ \frac{\cos{(M \Omega/2)}}{2\sqrt{M}}-\frac{\sin{(M \Omega/2)}}{2(\sqrt{M})^{3}} \right]=0,
\end{equation}
where $\Omega=\pi(\cos \hat{\phi}_{k}-\cos \phi_{k})$. It can be reformulated as follows.
\begin{equation}\label{tan}
\tan \left( \frac{\pi (\cos\hat{\phi}_k-\cos\phi_k)}{2} M \right)= \pi (\cos\hat{\phi}_k-\cos\phi_k)M,
\end{equation}
where its roots can be obtained by using Newton-Raphson method. $M$ which maximizes $c(M)$ can be represented as
\begin{equation}\label{tan_root}
\frac{\pi (\cos\hat{\phi}_k-\cos\phi_k)}{2} M= \pm 1.165.
\end{equation}
Since $M$ is constrained to a positive number, and the sign of $(\cos\hat{\phi}_k-\cos\phi_k)$ is unknown, $M_{k}$ can be represented as
\begin{equation}
{M}_{k}=\frac{2.330}{\pi \left| \cos\hat{\phi}_{k}-\cos{\phi_{k}} \right|}.
\end{equation}
\end{appendices}


\ifCLASSOPTIONcaptionsoff
  \newpage
\fi

\bibliographystyle{ieeetr}
\bibliography{reference}
\end{document}